\documentclass{svmult}

\usepackage{graphicx}    
\usepackage{a4}
\usepackage{amsmath}
\usepackage{amssymb}
\usepackage[T1]{fontenc}
\usepackage{graphicx}
\usepackage[varg]{txfonts}
\usepackage{natbib}
\usepackage{float}

\newcommand{\sil}{\lessapprox}
\newcommand{\mch}{$M_{\rm Chan}$~}
\begin{document}

\title*{Simulations of Turbulent Thermonuclear Burning in\\
  Type Ia Supernovae}
\titlerunning{Thermonuclear Supernovae}
\author{W. Hillebrandt, M. Reinecke,  W. Schmidt, F.~K. R{\"o}pke, C. Travaglio \inst{1}\and
J.~C.  Niemeyer\inst{2}}
\authorrunning{Thermonuclear Supernovae} 
\institute{Max-Planck-Institut f\"ur Astrophysik, Garching, Germany\\
\texttt{wfh@mpa-garching.mpg.de}
\and Institut f. Theor. Physik und Astrophysik, Univ. W\"urzburg,
Germany 
\texttt{ niemeyer@astro.uni-wuerzburg.de}}
%
%
\maketitle

\section{Summary}

Type Ia supernovae, i.e.\ stellar explosions which do not have hydrogen 
in their spectra, but intermediate-mass elements such as silicon,
calcium, cobalt, and iron, have recently received considerable 
attention because it appears that they can be used as ``standard
candles'' to measure cosmic distances out to billions of light years 
away from us. Observations of type Ia supernovae seem to indicate that 
we are living in a universe that started to accelerate its expansion 
when it was about half its present age. These conclusions rest
primarily on phenomenological models which, however, lack proper 
theoretical understanding, mainly because the explosion process,
initiated by thermonuclear fusion of carbon and oxygen into heavier 
elements, is difficult to simulate even on supercomputers. 

Here, we investigate a new way of modeling turbulent thermonuclear
deflagration fronts in white dwarfs undergoing a type Ia
supernova explosion. Our approach is based on a level set method
which treats the front as a mathematical discontinuity and
allows for full coupling between the front geometry and the flow
field. New results of the method applied to the problem of 
type Ia supernovae are obtained. It is shown that in
2-D with high spatial resolution and a physically motivated 
subgrid scale model for the nuclear flames numerically ``converged'' 
results can be obtained, but for most initial conditions the
stars do not explode. In contrast, simulations in 3-D,
do give the desired explosions and many of
their properties, such as the explosion energies, lightcurves and 
nucleosynthesis products, are in very good agreement 
with observed type Ia supernovae.   
\section{Introduction}
Numerical simulations of any kind of 
turbulent combustion have always been a challenge, and thermonuclear
supernova explosions are no exception to that rule. This is
mainly because of the large range of length scales involved. In type
Ia supernovae (SNe Ia), in particular, the length scales of relevant
physical processes range from 10$^{-3}$cm for the Kolmogorov-scale
to several 10${^7}$cm for typical convective motions.
In the currently favored scenario the explosion starts as a deflagration
near the center of the star. Rayleigh-Taylor unstable blobs of hot burnt
material are thought to rise and to lead to shear-induced turbulence
at their interface with the unburnt gas. This turbulence increases the
effective surface area of the flamelets and, thereby, the rate of fuel
consumption; the hope is that finally a fast deflagration 
might result, in agreement with phenomenological models of type
Ia explosions.

Despite considerable progress in the field of modeling turbulent
combustion for astrophysical flows the correct numerical 
representation of the thermonuclear deflagration front
has always been a weakness of the simulations. Methods used until recently 
were based on the reactive-diffusive flame model,
which artificially stretches the burning region over several grid zones 
to ensure an isotropic flame propagation speed. 
However, the soft transition from fuel to ashes stabilizes
the front against hydrodynamical instabilities on small length scales,
which in turn results in an underestimation of the flame surface area and
--~consequently~-- of the total energy generation rate. Moreover,
because nuclear fusion rates depend on temperature nearly
exponentially, one cannot use the zone-averaged values of the
temperature obtained this way to calculate the reaction kinetics.

The front tracking method used in this project cures most of these 
weaknesses. It is based on the so-called
\emph{level set technique} which was originally  
introduced by \cite{osher-sethian-88}. They used
the zero level set of a $n$-dimensional scalar function to represent
$(n-1)$-dimensional front geometries. Equations for the time evolution of
such a level set which is passively advected by a flow field are given in
\cite{sussman-etal-94}. The method has been extended to allow the tracking of
fronts propagating normal to themselves, e.g. deflagrations and detonations
\citep{smiljanovski-etal-97}. In contrast to the
artificial broadening of the flame in the reaction-diffusion-approach, this
algorithm is able to treat the front as an exact hydrodynamical discontinuity.
\section{The level set method}
\label{levelset_sect}
The central aspect of our front tracking method is the association
of the front geometry (a time-dependent set of points $\Gamma$)
with an isoline of a so-called level set function $G$:
\begin{equation}
  \Gamma:=\lbrace\vec r\ |\ G(\vec{r}) = 0\rbrace .
\end{equation}
Since $G$ is not completely determined by this equation, we can additionally
postulate that $G$ be negative in the unburnt and positive in the burnt
regions, and that $G$ be a ``smooth'' function, which is convenient from
a numerical point of view. This smoothness can be achieved, for example,
by the additional constraint that $|\vec{\nabla}G| = 1$
in the whole computational domain, with the exception of possible extrema
and kinks of
$G$. The ensemble of these conditions produces a $G$ which is a signed distance
function, i.e. the absolute value of $G$ at any point equals the minimal
front distance. The normal vector to the front is defined to
point towards the unburnt material.

The task is now to find an equation for the temporal evolution of $G$
such that the zero level set of $G$ behaves exactly
as the flame. Such an expression can be obtained by the consideration
that the total velocity of the front consists of two independent contributions:
it is advected by the fluid motions at a speed $\vec v$ and
it propagates normal to itself with a burning speed $s$.

Since for deflagration waves a velocity jump usually
occurs between the pre-front and post-front states, we must explicitly
specify which state $\vec v$ and $s$ refer to; traditionally, the values
for the unburnt state are chosen. Therefore, one obtains for the total
front motion
\begin{equation}
\label{Df}
  \vec D_f = \vec v_u + s_u \vec n.
\end{equation}
The total temporal derivative of $G$ at a point $P$ attached to the front
must vanish, since $G$ is, by definition, always 0 at the front:
\begin{equation}
  \frac{\text{d}G_P}{\text{d}t} = \frac{\partial G}{\partial t}
  + \vec \nabla G \cdot \dot {\vec x}_P = \frac{\partial G}{\partial t}
  + \vec D_f\cdot \vec\nabla G = 0
\end{equation}
This leads to the desired differential equation describing the time
evolution of $G$:
\begin{equation}
  \frac{\partial G}{\partial t} = - \vec D_f\cdot \vec\nabla G.
  \label{levprop}
\end{equation}

This equation, however, cannot be applied on the whole computational
domain, main\-ly because  using this equation everywhere will
in most cases destroy $G$'s distance function property.
Therefore additional measures must be taken in the regions away from the front
to ensure a ``well-behaved'' $|\vec\nabla G|$ \citep{reinecke-etal-99a}.

The situation is further complicated by the fact that the quantities
$\vec v_u$ and $s_u$ which are needed to determine $\vec D_f$ are not
readily available in the cells cut by the front. In a finite volume context,
these cells contain a mixture of pre- and post-front states instead.
Nevertheless one can assume that the conserved quantities (mass, momentum and
total energy) of the mixed state satisfy the following conditions:
\begin{alignat}{2}
  \overline{\rho}   &= \alpha \rho_u &&+ (1-\alpha) \rho_b \label{cons1}\\
  \overline{\rho \vec v} &= \alpha \rho_u \vec v_u
    &&+ (1-\alpha) \rho_b \vec v_b \label{cons2} \\
  \overline{\rho e} &= \alpha \rho_u e_u &&+ (1-\alpha) \rho_b e_b \label{cons3}
\end{alignat}
Here $\alpha$ denotes the volume fraction of the cell occupied by the unburnt
state.
In order to reconstruct the states before and behind the flame, a nonlinear
system consisting of the equations above, the Rankine-Hugoniot jump conditions
and a burning rate law must be solved. 

  \begin{figure}
  \centerline{\includegraphics[width=0.6\textwidth]{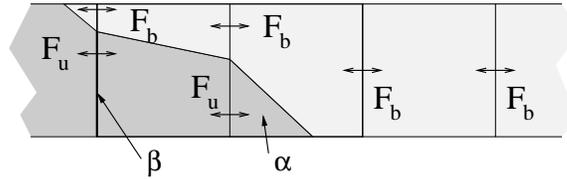}}
  \caption{
   Illustration of the basic principles of the level set method according
   to \cite{smiljanovski-etal-97}: The piecewise linear front cuts the mixed
   cells into burnt and unburnt parts. $\alpha$ is the unburnt volume fraction
   of a cell, $\beta$ is the unburnt area fraction of a cell interface. The
   fluxes $\vec F_u$ and $\vec F_b$ are calculated from the reconstructed
   states.}
  \label{flspl}
  \end{figure}

Having obtained the reconstructed pre- and post-front states in the mixed
cells, it is not only possible to determine $\vec D_f$, but also to separately
calculate the fluxes of burnt and unburnt material over the cell interfaces.
Consequently, the total flux over an interface can be expressed as a linear
combination of burnt and unburnt fluxes weighted by the unburnt interface area
fraction $\beta$ (see Fig. \ref{flspl}):
\begin{equation}
  \label{spliteq}
  \vec{\bar{F}} = \beta \vec F_u + (1-\beta)\vec F_b.
\end{equation}
\section{Implementation}
For our calculations, the front tracking algorithm was implemented as
an additional module for the hydrodynamics code PROMETHEUS
\citep{fryxell-etal-89}. Here we describe a simple implementation of
most of the ideas discussed in the previous section which we will call
``passive implementation''. It assumes that the $G$-function
is advected by the fluid motions and by burning and is only used to
determine the source terms for the reactive Euler equations. 
It must be noted that there exists no \emph{real} discontinuity between
fuel and ashes in this case; the transition is smeared out over
about three grid cells by the hydro-dynamical scheme, and the level set
only indicates where the thin flame front \emph{should} be. However,
the numerical flame is still considerably thinner than in the
reaction-diffusion approach.

A complete implementation contains in-cell-reconstruction and
flux-splitting as proposed by \cite{smiljanovski-etal-97} 
and outlined above. Therefore it
describes exactly the coupling between the flame and the hydrodynamic
flow. This generalized version of the code has 
been applied to hydrogen combustion in air by \cite{reinecke-etal-99a} and, 
very recently, also to thermonuclear fusion by \cite{roepke-etal-03}.

\subsection{$G$-Transport}
Since the front motion consists of two distinct contributions,
it is appropriate to use an operator splitting approach for the
time evolution of $G$. The advection term due to the fluid velocity
$\vec v_F$ reads in conservation form
\begin{equation}
\int_V\frac{\partial (\rho G)}{\partial t}d^3 r
+\oint_{\partial V} -\vec v_F \rho G d\vec f = 0
\end{equation}
\citep{mulder-etal-92}. This equation is identical to the advection equation
of a passive scalar, like the concentration of an inert chemical species.
Consequently, this contribution to the front propagation can be calculated
by PROMETHEUS directly.
The additional flame propagation due to burning is calculated at the end
of each time step and a re-initiali\-zation of $G$ is done in order to
keep it a signed distance function (see \citealt{reinecke-etal-99a}).

\subsection{Source terms}
After the update of the level set function in each time step, the change
of chemical composition and total energy due to burning is calculated in the
cells cut by the front. In order to obtain these values, the volume fraction
$\alpha$ occupied by the unburnt material is determined in those cells by the
following approach: from the value $G_{ij}$ and the two steepest gradients
of $G$ towards the front in $x$- and $y$-direction a first-order approximation
$\tilde G$ of the level set function is calculated; then the area fraction
of cell $ij$ where $\tilde G < 0$ can be found easily.
Based on these results, the new concentrations of fuel, ashes and energy are
obtained:
\begin{align}
X'_{\text{Ashes}} &= \text{max}(1-\alpha, X_{\text{Ashes}}) \label{newash} \\
X'_{\text{Fuel}} &= 1-X'_{\text{Ashes}} \\
e'_{\text{tot}} &= e_{\text{tot}} + q (X'_{\text{Ashes}}-X_{\text{Ashes}})
\end{align}
In principle this means that all fuel found behind the front is converted
to ashes and the appropriate amount of energy is released. The maximum operator
in eq. (\ref{newash}) ensures that no ``reverse burning'' (i.e. conversion from
ashes to fuel) takes place in the cases
where the average ash concentration is higher than the burnt volume fraction;
such a situation can occur in a few rare cases because of unavoidable
discretization errors of the numerical scheme.

\section{Turbulent nuclear burning}
\label{turb_burn}
The system of equations described so-far can be solved provided the
normal velocity of the burning front is known everywhere and at all
times. In our computations it is determined according to a flame-brush
model of \cite{niemeyer-hillebrandt-95a},
which we will briefly outline for convenience.

As was mentioned before, nuclear burning in degenerate dense matter
is believed to propagate on microscopic scales as a conductive
flame, wrinkled and stretched by local turbulence, but with essentially
the laminar velocity. Due to the very high Reynolds numbers macroscopic 
flows are highly turbulent and they interact with the flame,
in principle down to the Kolmogorov scale. This means
that all kinds of hydrodynamic instabilities feed energy into a
turbulent cascade, including the buoyancy-driven Rayleigh-Taylor
instability and the shear-driven Kel\-vin-Helm\-holtz instability.
Consequently, the picture that emerges is more that of a ``flame
brush'' spread over the entire turbulent regime rather than a wrinkled
flame surface. For such a flame brush, the relevant minimum length
scale is the so-called Gibson scale, defined as the lower bound for
the curvature radius of flame wrinkles caused by turbulent stress.
Thus, if the thermal diffusion scale is much smaller than the Gibson
scale (which is the case for the physical conditions of interest here)
small segments of the flame surface are unaffected by large scale
turbulence and behave as unperturbed laminar flames (``flamelets''). 
On the other hand side, since the Gibson scale is, at high densities,
several orders of magnitude smaller than the integral scale set by
the Rayleigh-Taylor eddies and many orders of magnitude larger than
the thermal diffusion scale, both transport and burning times are
determined by the eddy turnover times, and the effective velocity of
the burning front is independent of the laminar burning velocity.

A numerical realization of this general concept is presented in
\cite{niemeyer-hillebrandt-95a}. The basic assumption
was that wherever one finds turbulence this turbulence is fully
developed and homogeneous, i.e. the turbulent velocity fluctuations on
a length scale $l$ are given by the Kolmogorov law 
$v(l) = v(L) (l/L)^{1/3}$, 
where $L$ is the integral scale, assumed to be equal to the
Rayleigh-Taylor scale. Following the ideas outlined above, one can
also assume that the thickness  of the turbulent flame brush on the scale
$l$ is of the order of $l$ itself. With these two assumptions and the
definition of the Gibson scale one finds
 for $l_{\mathrm{gibs}} \sil l \sil L \simeq \lambda _{\mathrm{RT}}$
\begin{equation} 
\label{vtur}
v(l) \simeq s_{\mathrm{t}}(l) \simeq s_{\mathrm{t}}(l_{\mathrm{gibs}}) \biggl(\frac{l}{l_{\mathrm{gibs}}} 
\biggr)^{1/3}  
\end{equation}
and $d_t(l) \simeq l$, 
where $v(l_{\mathrm{gibs}}) = s_{\mathrm{lam}}$ defines $l_{\mathrm{gibs}}$,
$s_{\mathrm{lam}}$ is the
laminar burning speed and $s_{\mathrm{t}}(l)$ is the
turbulent flame velocity on the scale $l$.

In a second step this model of turbulent combustion is coupled to our
finite volume hydro scheme.  Since in every finite volume scheme
scales smaller than the grid size cannot be resolved, we express
$l_{\mathrm{gibs}}$ in terms of the grid size $\Delta$, the (unresolved)
turbulent kinetic energy per unit mass, $k_{\mathrm{sgs}}$, and the
laminar burning velocity:
\begin{equation}
\label{gibs}
l_{\mathrm{gibs}} = \Delta \biggl( \frac{u_l^2}{2k_{\mathrm{sgs}}} \biggr) ^{3/2}.
\end{equation}
Here $k_{\mathrm{sgs}}$ is determined from a subgrid scale model
\citep{clement-93,niemeyer-hillebrandt-95a} and, finally, the effective
turbulent velocity of
the flame brush on scale $\Delta$ is given by 
\begin{equation}
\label{vburn}
s_{\mathrm{t}}(\Delta) = \operatorname{max} ( s_{\mathrm{lam}}, v(\Delta), v_{\mathrm{RT}}),
\end{equation}
with $v(\Delta) = \sqrt {2k_{\mathrm{sgs}}}$ and $v_{\mathrm{RT}} \propto \sqrt {g \Delta}$,
where $g$ is the local gravitational acceleration.

\section{Application to the supernova problem}
\label{applict}

Different series of simulations were performed to check the numerical
reliability of the employed models and to compare two- and three-dimensional
explosions.

\begin{figure}
  \centerline{\includegraphics[width=0.7\textwidth]{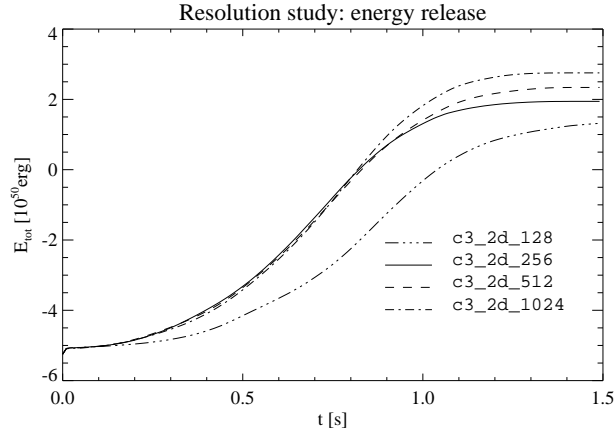}}
  \caption{Time evolution of the total energy for identical initial conditions
    at different resolutions. While model c3\_2d\_128 is clearly under-resolved,
    the other simulations agree very well, at least in the early and
    intermediate stages.}
  \label{e_compare}
\end{figure}

\subsection{Resolution study}
A crucial test for the validity of the models for the unresolved scales
(in this case the flame and subgrid models) is to check the dependence of
integral quantities, like the total energy release of the explosion, on the
numerical grid resolution. Ideally, there should be no such dependence,
indicating that all effects on unresolved scales are accurately modeled.

Figure \ref{e_compare} shows the energy evolution of a centrally ignited white
dwarf. The only difference between the simulations is the central grid
resolution,
which ranges from $2\cdot10^6$cm (model c3\_2d\_128) down to $2.5\cdot10^5$cm
(model c3\_2d\_1024). Model c3\_2d\_128 is obviously under-resolved, but the
results of the other calculations are in good agreement, with exception of
the last stages, where the flame enters strongly non-uniform regions of the
grid.

So far, this kind of parameter study could only be performed in two dimensions,
because of the prohibitive cost of very highly resolved 3D simulations.
Nevertheless the results suggest that a resolution of $\approx10^6$cm should
yield acceptable accuracy also in three dimensions. 

\begin{figure}
  \centerline{\includegraphics[width=0.7\textwidth]{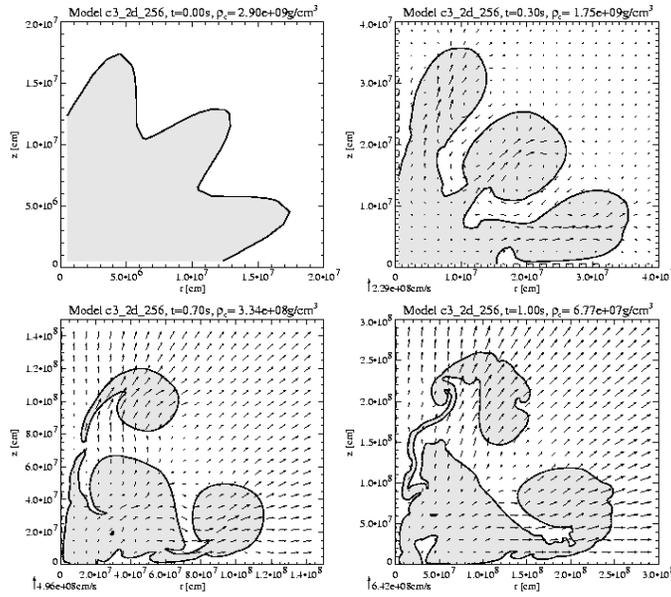}}
  \caption{Burning front geometry evolution of model c3\_2d\_256.}
  \label{front2d}
\end{figure}

\begin{figure}[tbp]
  \centerline{\includegraphics[height=0.6\textheight]{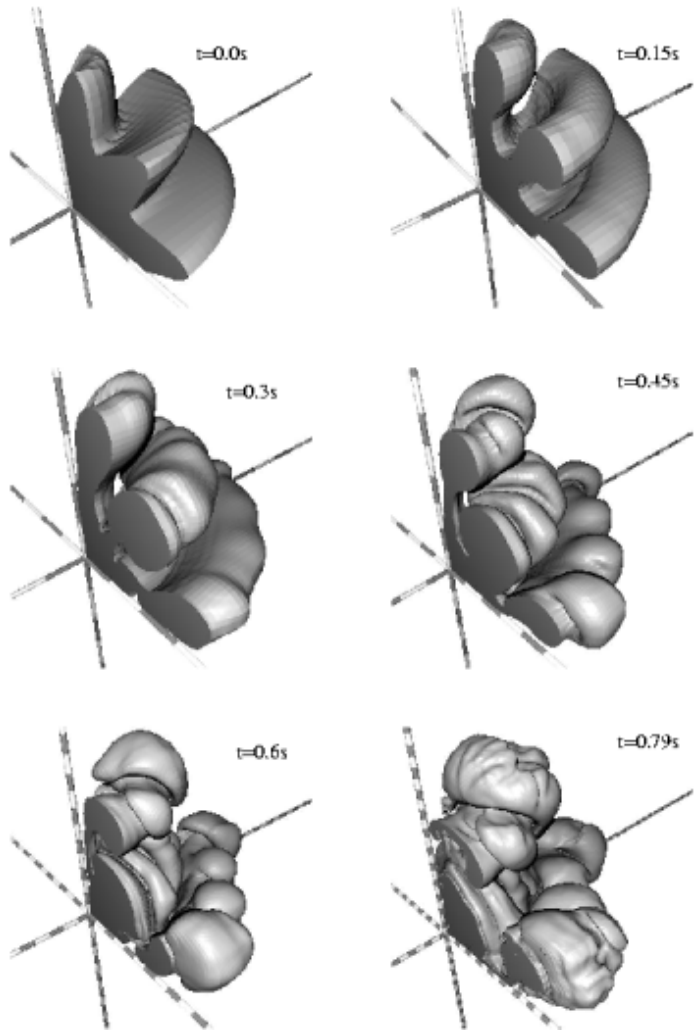}}
  \caption{Burning front geometry evolution of model c3\_3d\_256.
           One ring on the axes corresponds to $10^7$cm.}
  \label{front3d}
\end{figure}

\begin{figure}
  \centerline{\includegraphics[width=0.7\textwidth]{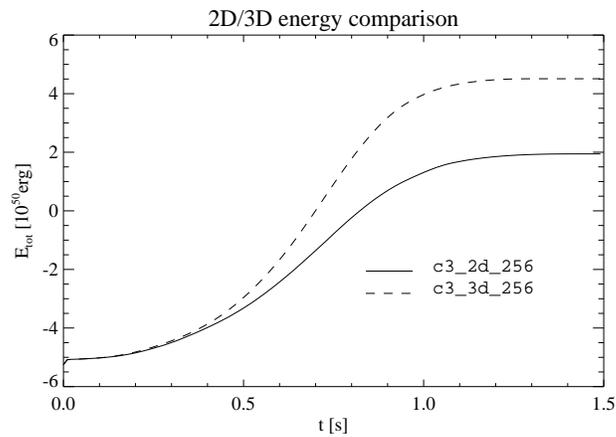}}
  \caption{Comparison of the energy evolution for identical initial
           conditions in two and three dimensions.}
  \label{comp2d3d}
\end{figure}

\subsection{Comparison of 2D and 3D simulations}
In order to investigate the fundamental differences between two- and
three-dimensional simulations, a 2D and a 3D model with identical initial
conditions and resolution was calculated.
Figures \ref{front2d} and \ref{front3d} show
snapshots of the flame geometry at various explosion stages; the energy
evolution of both models is compared in figure \ref{comp2d3d}.

It is evident that both simulations evolve nearly identically during the first
few tenths of a second, as was expected. This is a strong hint that no errors
were introduced into the code during the enhancement of the numerical models
to three dimensions. At later times, however, the 3D calculation develops
instabilities in the azimuthal direction, which could not form in 2D because
of the assumed axial symmetry. As a consequence the total burning surface
and the energy generation rate is increased, resulting in a higher overall
energy release.

\subsection{The effect of different initial conditions}

In our approach, the initial white dwarf model (composition, 
central density, and velocity structure), as well as assumptions 
about the location, size and shape of the flame surface as it first 
forms fully determine the simulation results. At present, we have not
changed the properties of the white dwarf but we concentrate on 
variations of the latter. In this context, the simultaneous runaway at 
several different spots in the central region of the progenitor star 
is of particular interest, a plausible ignition scenario suggested 
by \cite{garcia-woosley-95}.

\begin{figure}
\begin{tabular}{cc}
{\includegraphics[width=0.48\textwidth]{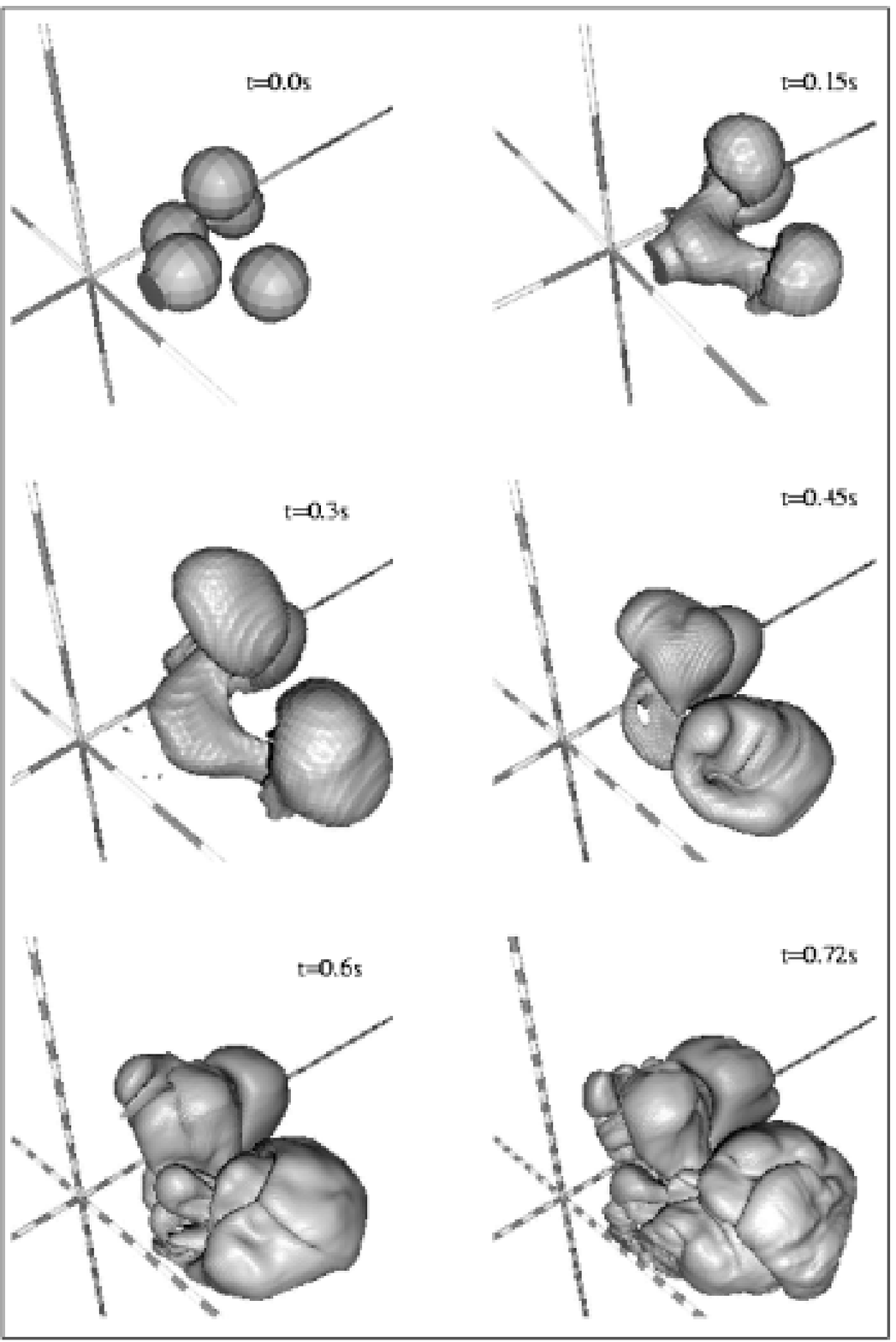}}
&
{\includegraphics[width=0.48\textwidth]{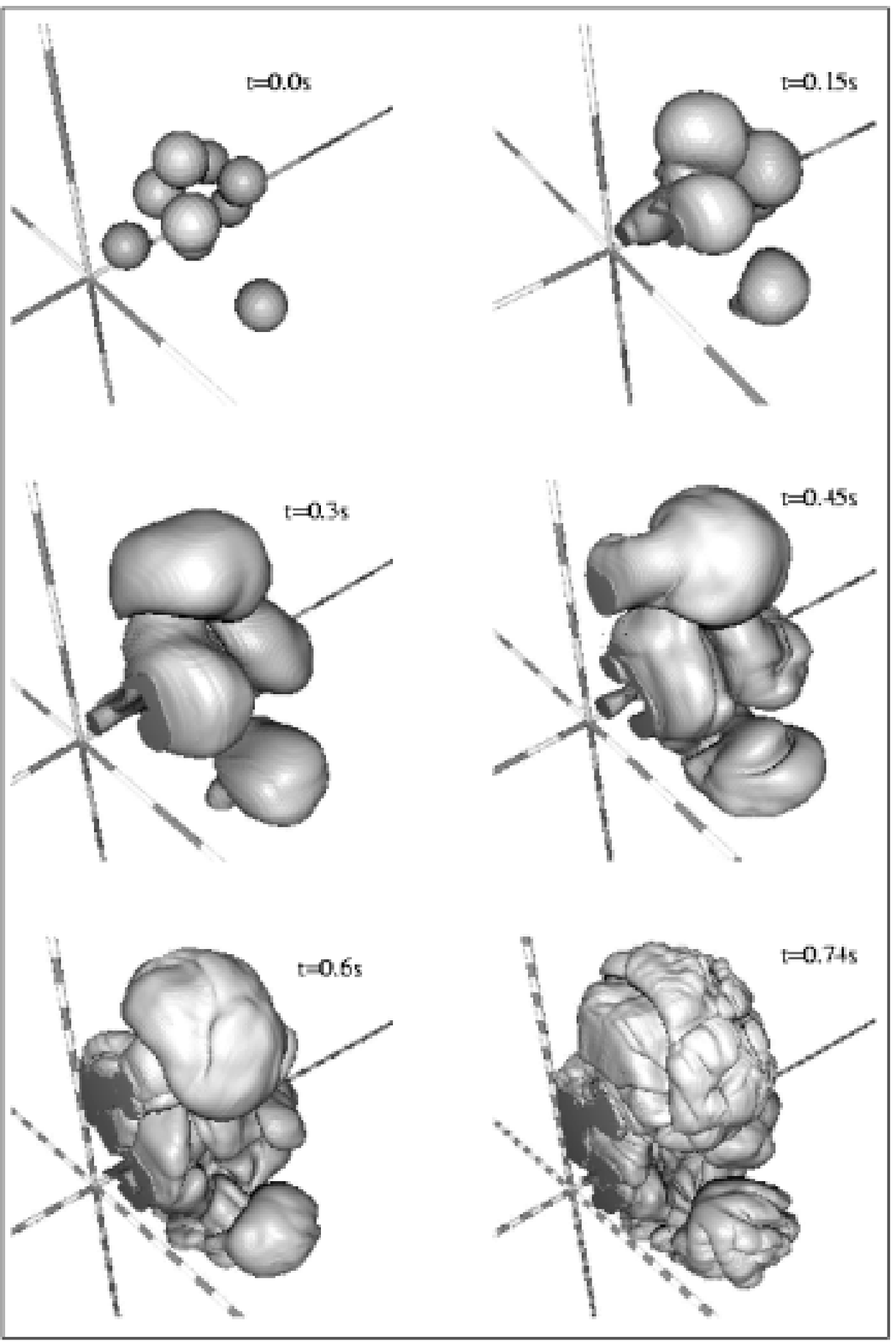}}
\end{tabular}
\caption{Left panel: Snapshots of the flame front for a scenario 
with 5 ignition spots in 3D. The fast merging between the leading 
and trailing bubbles and the rising of the entire burning region 
is clearly visible. One ring on the coordinate axes corresponds 
to $10^7$cm. Right panel: The same as before, but for a high 
resolution model with 9 ignition spots.}
\end{figure}

\begin{figure} 
\centerline{\includegraphics[width=0.7\textwidth]{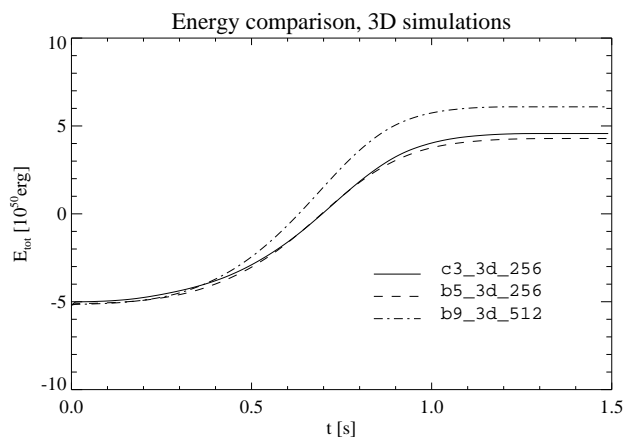}}
\caption{Energy evolution of these 
three-dimensional explosion models (dashed and dashed-dotted). For
comparison we also show the centrally ignited (``three fingers'') model of
\cite{reinecke-etal-02a} (solid line) discussed earlier.}
\label{multipoint}
\end{figure}

One simulation was carried out on a grid of 256$^3$ cells with a
central resolution of 10$^6$\,cm and contained five
bubbles with a radius of 3$\cdot$10$^6$\,cm, which were distributed randomly
in the simulated octant within 1.6$\cdot$10$^7$\,cm of the star's center.
In an attempt to reduce the initially burned mass as much as possible without
sacrificing too much flame surface, a very highly resolved second model 
was constructed. It contains nine randomly distributed, non-overlapping
bubbles with a radius of 2$\cdot$10$^6$\,cm within 1.6$\cdot$10$^7$\,cm 
of the white dwarf's center. To properly represent these very small
bubbles, the cell size was reduced to $\Delta=$5$\cdot$10$^5$\,cm, so 
that a total grid size of $512^3$ cells was required.

During the first 0.5 seconds, the three models
are nearly indistinguishable as far as the total energy is
concerned (see Fig.\ \ref{multipoint}), which at first glance appears
somewhat surprising, given the quite different initial conditions.
A closer look at the energy generation rate actually reveals noticeable
differences in the intensity of thermonuclear burning for the 
simulations, but since the total flame surface is initially very small,
these differences have no visible impact on the integrated curve in the
early stages.

However, after about 0.5 seconds, when fast energy generation sets in, 
the nine-bubble model burns more vigorously due to its larger surface 
and therefore reaches a higher final energy level. Fig.\ \ref{multipoint}
also  shows that the centrally ignited model (c3\_3d\_256) 
is almost identical to the off-center model b5\_3d\_256 with regard to 
the explosion energetics. But, obviously, the scatter in the final 
energies due to different initial conditions appears to be small.
Moreover, all models explode with an explosion energy in the
range of what is observed.

\section{Predictions for observable quantities}

In this Section we present a few preliminary results for various
quantities which could, in principle, be observed and which
therefore can serve as tests for the models.

\subsection{Lightcurves}

The most direct test of explosion models is provided by observed
lightcurves and spectra. According to ``Arnett's Law'' lightcurves
measure mostly the amount and spatial distribution of radioactive
$^{56}$Ni in type Ia supernovae, and spectra measure the chemical
composition in real and velocity space.

\begin{figure}
\centerline{\includegraphics[width=0.7\textwidth]{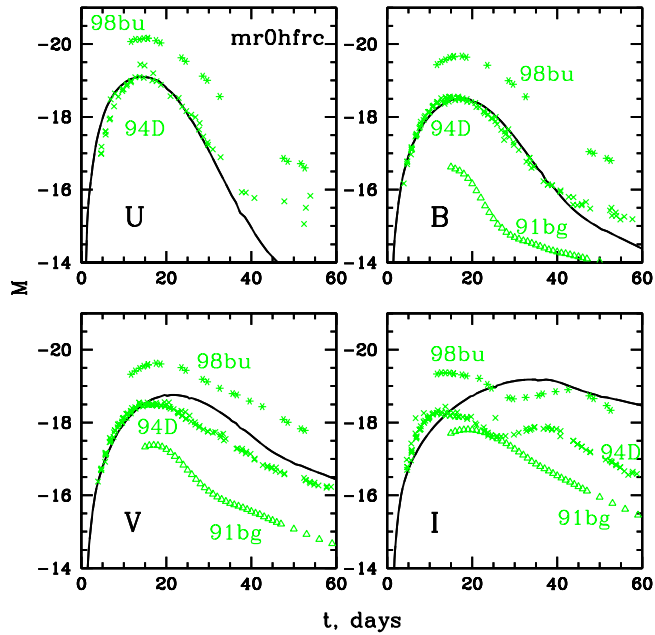}}
\caption{UBVI-colour lightcurves predicted by a centrally ignited 3D
model (solid lines) in comparison with obseved data for a bright
(98bu), a ``normal'' (94D), and a subluminous (91bg) supernova.}
\label{lightcurves}
\end{figure}

\cite{sorokina-blinnikov-03} have used the results of one of our
centrally ignited 3D-model, averaged over spherical shells,
to compute colour lightcurves in the UBVI-bands. Their code
assumes LTE radiation transport and loses reliability at later
times (about 4 weeks after maximum) when the supernova enters
the nebular phase. Also, this assumption and the fact that the
opacity is not well determined at longer wavelength make I-lightcurves
less accurate. Keeping this in mind, the lightcurves shown in
Fig.\ \ref{lightcurves} look very promising. The main reason for the
good agreement between the model and SN 1994D is the presence of radioactive 
Ni in outer layers of the supernova model at high velocities 
which is not be predicted by spherical models.

\subsection{Elemental and isotopic abundances}

A summary of the abundances obtained for all 3D models is given in Table 1.
Here ``Mg'' (as in Fig.\ \ref{lightcurves}) stands for
intermediate-mass nuclei, and ``Ni'' for the iron-group. 
In addition, the total energy liberated by nuclear burning is given.
Since the binding energy of the white dwarf was about 5$\cdot$10$^{50}$
erg, all models do explode. Typically one expects that around
80\% of iron-group nuclei are originally present as $^{56}$Ni
bringing our results well into the range of observed Ni-masses. 
This success of 
the models was obtained without introducing any non-physical 
parameters, but just on the basis of a physical and numerical model 
of subsonic turbulent combustion. We also stress that our models 
give clear evidence that the often postulated deflagration-detonation 
transition is not needed to produce sufficiently powerful explosions.

 \begin{table}[htbp]
  \centerline{
    \begin{tabular}{|l|c|c|c|}
    \hline
    model name & $m_{\text{Mg}}$ [$M_\odot$]&$m_{\text{Ni}}$ [$M_\odot$]&$E_{\text{nuc}}$ [$10^{50}$\,erg\vphantom{\raisebox{2pt}{\large A}}] \\
    \hline
    c3\_3d\_256 & 0.177 & 0.526 & 9.76 \\ \hline
    b5\_3d\_256 & 0.180 & 0.506 & 9.47 \\ \hline
    b9\_3d\_512 & 0.190 & 0.616 & 11.26\phantom{0} \\ \hline
    \end{tabular}
  }
  \caption{Overview over element production and energy release of all
    discussed supernova simulations}
  \label{burntable}
\end{table}

\begin{figure}
\centerline{\includegraphics[width=0.70\textwidth]{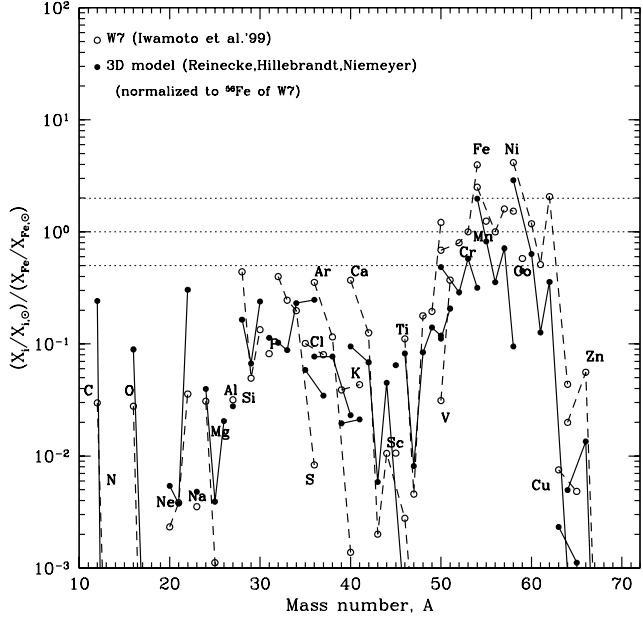}}
\caption{Isotopic abundances obtained for off-center ignited 3D
model.}
\label{abundances}
\end{figure}

Finally, we have ``post-processed'' one of our models in order to 
see whether or not also reasonable isotopic abundances are
obtained. The results, shown in Fig.\ \ref{abundances}, are 
preliminary and should be considered with care. However, it is
obvious that, with a few exceptions, also isotopic abundances
are within the expected range and do not differ too much from those 
computed by means of phenomenological models, i.e. ``W7''. Exceptions  
include the high abundance
of (unburned) C and O, and the overproduction of $^{48,50}$Ti,
$^{54}$Fe, and $^{58}$Ni.   

\section{Supplementary studies}

In order to validate our approach to model the effects on unresolved
scales, two studies were carried out. One tests several subgrid scale
(SGS) models under the physical conditions encountered in SN Ia
explosions and the other concerns the flame stability on small scales,
where the turbulent cascade does not dominate the flame evolution.

\begin{figure}
  \centerline{\includegraphics[width=0.95\textwidth]{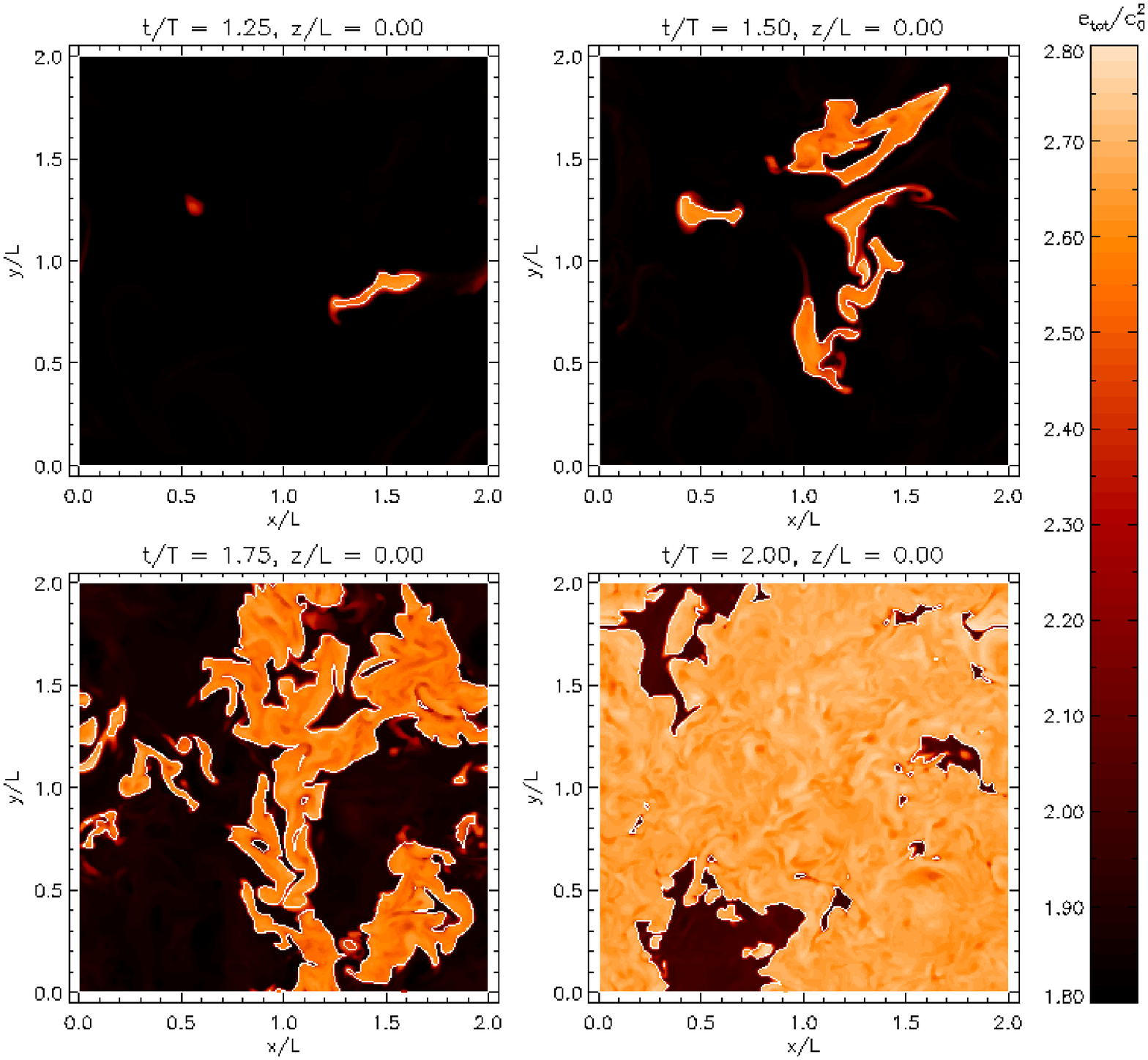}}
  \caption{Evolution of turbulent
    thermonuclear burning in a cubic domain subject to periodic
    boundary conditions. The panels show 2D contour sections of the
    normalized specific total energy, $e_{\mathrm{tot}}/c_{0}^{2}$.
    The mean mass density is $\rho_{0}\approx
    2.903\cdot10^{8}\,\mathrm{g\,cm^{-3}}$ and the initial sound speed
    $c_{0}\approx 6.595\cdot 10^{8}\,\mathrm{cm\,s^{-1}}$. Turbulence
    is produced artificially by a stochastic solenoidal force field.
    The bright regions of high specific energy contain burned
    material. The flame front is indicated by the thin white lines
    corresponding to the zero level set. }
  \label{les216_energy}
\end{figure}

\begin{figure}
  \begin{center}
    \includegraphics[width=1.0\linewidth]{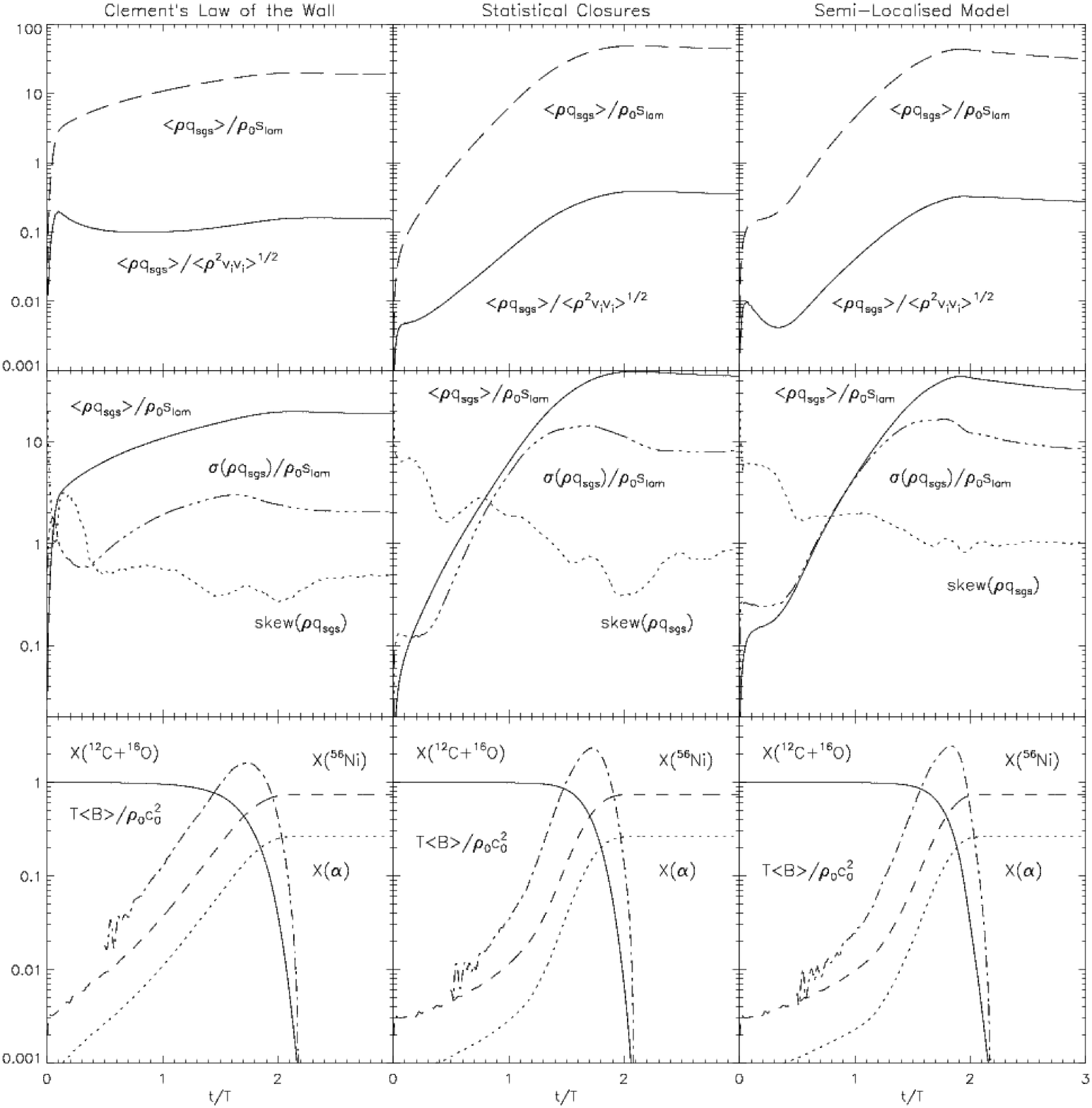}
    \caption{ Comparison of the evolution of SGS turbulence and
      thermonuclear burning for LES with three different variants of
      the SGS turbulence energy model.  In one simulation, Clement's
      law of the wall was used to calculate the closure parameters,
      the second LES was computed with the constant parameters
       and for the third simulation, a
      semi-localized model was applied. In the top panels, the ratio
      of the mean SGS turbulence velocity to the RMS velocity and the
      laminar burning speed, respectively, is shown as function of
      normalized time $t/T$. In the middle row of panels, the first
      three statistical moments of the mass-weighted SGS turbulence
      velocity and, in the bottom panels, the time evolution of fuel,
      burning products and the mean energy generation rate are
      plotted. }
    \label{fg:les144_clemt}
  \end{center}
\end{figure}

\subsection{Large eddy simulations of turbulent deflagration}

Subgrid scale models were tested in large eddy simulations
(LES) of thermonuclear burning in a turbulent flow. In these
simulations, turbulence is driven by an artificial stochastic force
field and the computational domain is cubic with periodic boundary
conditions.  The term \emph{large eddy simulation} indicates that not
all dynamic scales are numerically resolved. Compared to
simulations of supernova explosions, there are two additional
differences: Firstly, the periodic boundary conditions enforce a
constant average mass density. Hence, there is no explosion in this
setting. The burning process proceeds in the form of a
\emph{percolation process} as gradually all fuel is consumed and the
system approaches a state in which degenerate C+O matter is replaced
by ${^{56}}$Ni and $\alpha$ particles in statistical nuclear
equilibrium.  Secondly, gravity is not included. Since the integral
length scale is of the order $10^{5}\,\mathrm{cm}$, this is a sensible
approximation. In consequence, we focus on the interaction of the
deflagration with purely \emph{hydrodynamical} turbulence for which
the similarity theory of Kolmogorov applies.  Whether SGS models which
do not account for effects induced by gravity are applicable to
thermonuclear supernova explosions is still controversial. However,
there are simple scaling arguments in favor of this
conjecture. The single most important result emerging from our
studies is that the evolution of the burning process is significantly
affected by the SGS model in use.  This underlines the importance of
choosing a faithful SGS model and certainly bears consequences on the
simulation of deflagration in SNe Ia.

Our approach of determining the turbulent flame speed as outlined in
Section~\ref{turb_burn} is based upon a dynamical equation for the SGS
\emph{turbulence energy}
$k_{\mathrm{sgs}}=\frac{1}{2}q_{\mathrm{sgs}}^{2}$ which can be casted
into the following equation for the turbulence velocity
$q_{\mathrm{sgs}}$:
\begin{equation}
  \label{eq:sgs_velocity_cld}
  \frac{\mathrm{D}}{\mathrm{D}t}q_{\mathrm{sgs}} - \frac{1}{\rho}
  \nabla\cdot\left(\rho\ell_{\kappa}q_{\mathrm{sgs}}
                   \nabla q_{\mathrm{sgs}}\right) -
  \ell_{\kappa}|\nabla q_{\mathrm{sgs}}|^{2} =
  \ell_{\nu}|S^{\ast}|^{2} -
  \left(\frac{1}{3}+\frac{C_{\lambda}}{2}\right)q_{\mathrm{sgs}}d -
  \frac{q_{\mathrm{sgs}}^{2}}{\ell_{\epsilon}}
\end{equation}
The operator $\frac{\mathrm{D}}{\mathrm{D}t}$ is the \emph{Lagrangian}
time derivative $\frac{\partial}{\partial t}+\vec v\cdot\nabla$.  In
this equation, several heuristic approximations, so-called
\emph{closures}, are incorporated.  Associated with these closures are
the length scales $\ell_{\nu}=C_{\nu}\Delta_{\mathrm{eff}}/\sqrt{2}$,
$\ell_{\epsilon}=2\sqrt{2}\Delta_{\mathrm{eff}}/C_{\epsilon}$ and
$\ell_{\kappa}=C_{\kappa}\Delta_{\mathrm{eff}}/\sqrt{2}$.
$C_{\lambda}$ accounts for pressure effects in a compressible
fluid. The closure parameters $C_{\nu}$, $C_{\epsilon}$, $C_{\kappa}$
and $C_{\lambda}$ are \emph{a priori} unknown.  In the case of
stationary isotropic turbulence in an incompressible fluid, values can
be derived from analytic theories of turbulence.
Particular examples are $C_{\nu}\approx 0.054$, $C_{\epsilon}\approx
1.0$ and $C_{\kappa}\approx 0.1$ (cf.\ \cite{Sagaut}, Section 4.3) and
$C_{\lambda}\approx-0.2$ \citep{FurTab97}.  $\Delta_{\mathrm{eff}}$ is
the \emph{effective length scale} of the finite-volume scheme, in this
case, the piece-wise parabolic method. Usually, this scale is set
equal to the size $\Delta$ of the grid cells. However, from the energy
spectra of direct numerical simulations (DNS) of forced isotropic
turbulence, we concluded that is more appropriate to set
$\Delta_{\mathrm{eff}}=\beta\Delta$, where $\beta$ is in the range
$1.6\ldots 1.8$ depending on the Mach number. The factor $\beta$
accounts for the smoothing effect of the numerical scheme on top of
the discretization.

We have also determined statistical values of the closure parameters
from DNS data.  For subsonic flows, $C_{\nu}=0.06$,
$C_{\epsilon}=0.48$ and $C_{\kappa}=0.36$ appear to be representative
values. The outcome of a LES of turbulent deflagration in a cube of
$216^{3}$ cells with these parameters is illustrated in
Figure~\ref{les216_energy}. Burning is ignited in eight small
spherical zones at the beginning of the simulation and the
fluid is set into motion by stochastic forcing.  Turbulence is
produced on a time scale $T=L/V$, where $L=2.16\cdot
10^{5}\,\mathrm{cm}$ is the integral length scale and
$V=100s_{\mathrm{lam}}\approx 9.78\cdot 10^{7}\,\mathrm{cm\,s^{-1}}$ is
the characteristic velocity of the flow. One can think of $L$ being
the typical size of the largest vortices generated by the stochastic force field
and $T$ is the associated autocorrelation time or, figuratively, the turn-over
time. Initially, the burning zones are expanding very slowly as the
flame propagation speed mostly equals the laminar speed. After roughly
one turn-over time has elapsed, SGS turbulence becomes increasingly
space filling and enhances the flame propagation speed. At the same time,
the folding and stretching of the flame front by the resolved flow
greatly increases the surface and, thus, the rate of fuel consumption.
These two effects appreciably accelerate the burning process as one
can see from the evolution of the energy shown in
Figure~\ref{les216_energy}. At time $t\gtrsim 2T$, most of the
material has already been burnt.

Constant closure parameters are a sensible choice for fully developed
isotropic turbulence. However, significant deviations are expected for
transient, intermittent or anisotropic flows. The deflagration in the
cube is certainly transient and, naturally, the physical conditions in
the vicinity of the flame front are anisotropic. In the case of a
supernova explosion, these qualities are even more pronounced. For
this reason, a more sophisticated method of determining the closure
parameters is called for. \cite{clement-93} found for stellar structure
modeling \emph{ad hoc} rules to adjust the parameters $C_{\nu}$ and
$C_{\epsilon}$ in order to account for compressibility in stratified
media.  Since steep density gradients affect turbulence more or less
like a wall, we shall refer to the rules proposed by Clement as ``law
of the wall''.  This method has been applied to SGS modeling in the
simulations of thermonuclear supernovae as well. In comparison to the SGS
turbulence energy model with constant parameters, Clement's law of the
wall produces markedly different results as one can seen from the
LES statistics plotted in Figure~\ref{fg:les144_clemt}. The mean of
$q_{\mathrm{sgs}}$ increases initially very rapidly and then settles
at an almost constant level.  Correspondingly, the average rate of
energy release per unit volume due to thermonuclear burning, $\langle B\rangle$,
increases more gradually as opposed to steep rise and high peak found
for the SGS model with constant parameters.

An entirely different method utilizes the \emph{filtering approach}
introduced by \cite{Germano92}. Smoothing the resolved velocity field
with a filter of characteristic length larger than
$\Delta_{\mathrm{eff}}$ and invoking self-similarity assumptions,
equations can be derived which yield closure parameters varying in
space and time corresponding to the local structure and evolution of
the flow. \citep{Pio93,LiuMen94,GhoLund95,MeneKatz00}.  This
so-called \emph{dynamical procedure} works particularly well for the
production parameter $C_{\nu}$.  In the case of dissipation and diffusion,
however, the method appears to be inadequate because of deficiencies in the
respective closures. Therefore, we adopted a semi-localized approach.  
Results from a LES using this method, are shown in the
panels on the very right of Figure~\ref{fg:les144_clemt}. The
differences compared to the LES with constant parameters are not
striking but Clement's method is clearly dismissed. The
prediction of the latter that SGS turbulence is produced right at the
beginning in \emph{advance} of small-scale structure developing in the
resolved flow is physically unreasonable.  Apart from that, once
turbulence has developed in a particular spatial region, it becomes
increasingly isotropic on decreasing length scales even in more
complex flows. Thus, the remarkable similarity of the evolution of the
burning process in the LES with constant and dynamical parameters,
respectively.  However, the choice of appropriated values for the
closure parameters is notoriously difficult, particularly, when it
comes to the supernova explosion scenario in which turbulence is
generated due to buoyancy effects originating from the density
contrast between burnt and unburnt material. In conclusion, the
semi-localized model is likely to be the preferable SGS model for the
application to SNe Ia simulations after all.

\subsection{Investigation of the cellular burning regime}

\begin{figure}[t]
\centerline{
\includegraphics[width=0.9 \textwidth]{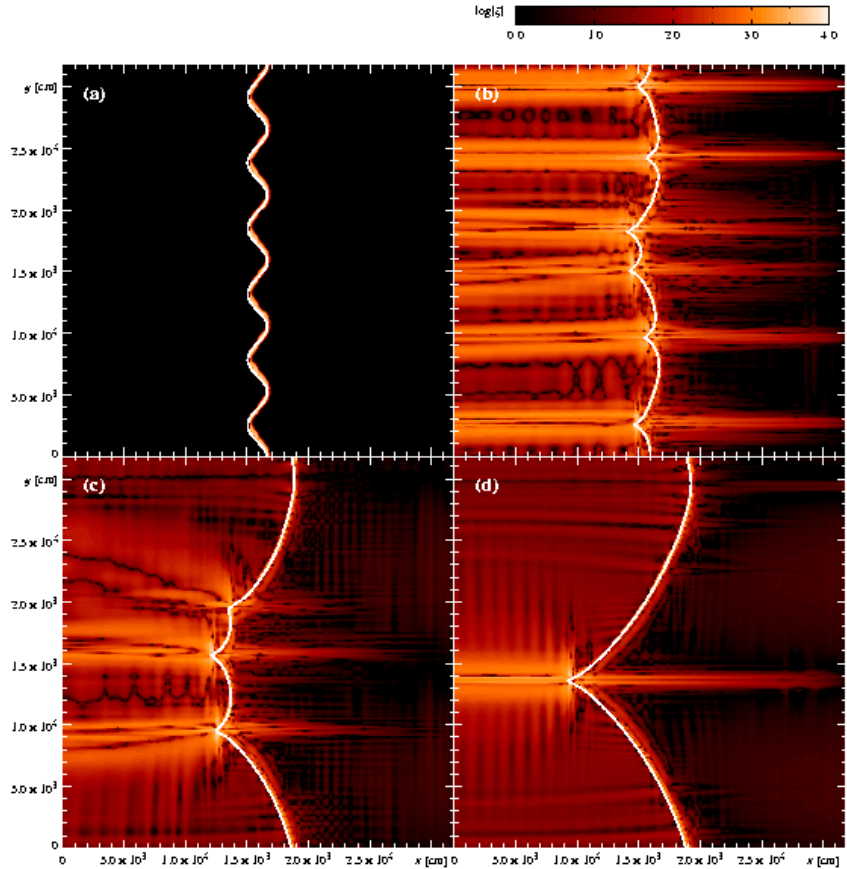}}
\caption{Flame propagation into quiescent fuel at $\rho_u = 5 \times
  10^7 \,\mathrm{g} \,\mathrm{cm}^{-3}$, resolution: $200 \times 200$ cells,
  snapshots taken at \textbf{(a)} 0, \textbf{(b)} 7.5, \textbf{(c)}
  15, and \textbf{(d)} 30 growth times
  $\tau_\mathrm{LD}$ of a
  perturbation with $\lambda=3.2 \times 10^4 \,
  \mathrm{cm}$. Color-coded is the vorticity of the flow field. (The
  figure is also included in the color plates of this volume.)
  \label{quif_fig}}
\end{figure}

Below the Gibson scale, flame propagation is not dominated by the
turbulent cascade, since here the flame burns faster through turbulent
eddies than these can deform it.
On those scales flame evolution is determined by the
(hydrodynamical) Landau-Darrieus (LD)
instability \citep{landau-44} and its counteracting nonlinear
stabilization \citep{zeldovich-66}. In
case of terrestrial flames this
stabilization leads to a cellular
steady-state pattern of the flame front giving rise to the
\emph{cellular burning regime}. After \citet{niemeyer-hillebrandt-95b}
have shown by means of hydrodynamical simulation that the LD
instability acts under conditions of SNe Ia, we could confirm that
also the
cellular stabilization of the flame 
front holds for thermonuclear flames in white dwarfs. This was
achieved in a series of 2D simulations. 
In order to reproduce hydrodynamical effects like the LD instability
it turned out to be essential to apply the \emph{complete
  implementation} described in Sect.~\ref{levelset_sect} (see \citealt{roepke-etal-03}).
The
stability of this cellular flame shape was tested for various fuel
densities and for interaction of the flame with vortical flows. Two
examples are given in the following. 

Figure \ref{quif_fig} shows the propagation of an initially perturbed
flame into quiescent fuel. The initial perturbations grow and in the
nonlinear regime the flame exhibits a cellular structure. For the given
setup, however, the cellular pattern of the same wavelength as the
initial perturbation is not a stable solution. The snapshots at later
times (Fig.~\ref{quif_fig}b,c) illustrate the ``merging'' of cells
until the steady-state structure of a single domain-filling cusp has
formed. In case of higher numerical resolution, this fundamental flame
structure may be superposed by a short-wavelength cellular pattern,
which is advected toward the cusp and does not lead to a break-up of
the domain-filling cell. This evolution is well in agreement with theoretical
predictions. Simulations with different fuel densities $\rho_u$ led to
similar results. In the range of $\rho_u = 1\times 10^7 \,\mathrm{g}
\,\mathrm{cm}^{-3} \ldots 1\times 10^9 \,\mathrm{g}
\,\mathrm{cm}^{-3}$ no significant flame destabilization could be
observed. 

\begin{figure}[t]
\centerline{
\includegraphics[width=0.9 \textwidth]{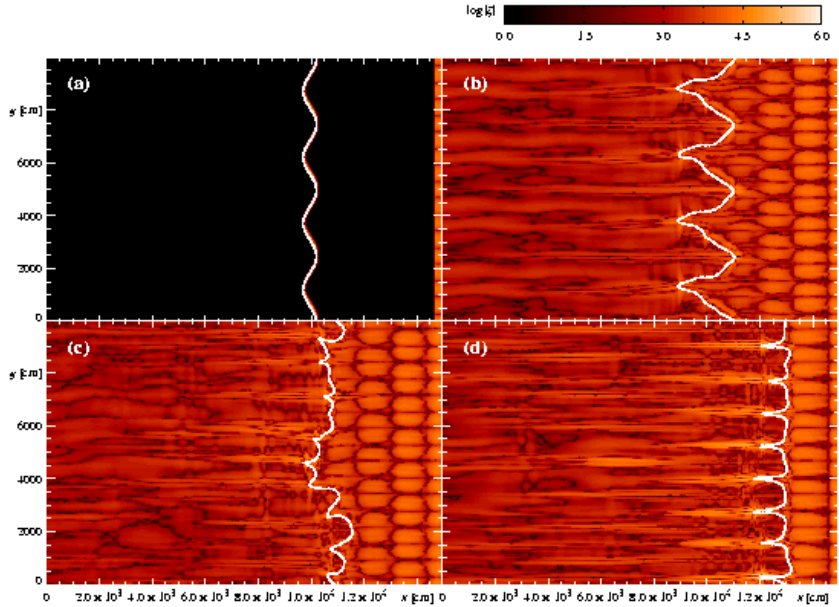}}
\caption{Flame propagation into vortical fuel at $\rho_u = 5 \times
  10^7 \,\mathrm{g} \,\mathrm{cm}^{-3}$; velocity fluctuations at the
  right boundary: $v'/u_\mathrm{lam} = 2.5$; resolution: $300 \times 200$ cells;
  snapshots taken at \textbf{(a)} $0 \, \mathrm{s}$, \textbf{(b)} $8.0
  \times 10^{-3} \, \mathrm{s}$, \textbf{(c)} $1.6
  \times 10^{-2} \, \mathrm{s}$, and \textbf{(d)} $4.8
  \times 10^{-2} \, \mathrm{s}$.  Color-coded is the vorticity of the
  flow field. (The
  figure is also included in the color plates of this volume.)
  \label{vorf_fig}}
\end{figure}

Flame interaction with a vortical fuel field is shown in
Fig.~\ref{vorf_fig}. This is motivated by relic turbulent motions from
the eddy cascade around the Gibson scale and from pre-ignition
convective motions in the white dwarf.
A parameter study with different fuel densities
and various strengths of the velocity fluctuations in the fuel flow
led to the result that a vortical flow of sufficient strength can
break up the cellular pattern of the flame. In this case, however, no
drastic self-turbulization of the flame but rather a smooth adaptation
of the flame structure to the imprinted flow was observed.

These results corroborate the assumption of large
scale models, that the generation of turbulence is dominated by
large-scale effects.
In the cellular burning regime, the flame surface is enlarged compared
to a planar configuration. This leads to an additional acceleration of
the flame.
As a result of our study we suggest an increased lower cut-off of the flame
velocity (cf.~eq.~(\ref{vburn})) instead of $u_\mathrm{lam}$ depending
on background turbulence.

\section{Conclusion}

In this project, we have provided a new method to model the 
physics of thermonuclear combustion in degenerate dense matter 
of white dwarfs stars consisting of carbon and oxygen.
Because not all relevant length-scales of this problem
can be numerically resolved a numerical model to describe 
deflagration fronts with a reaction zone much thinner then the 
cells of the computational grid was presented. This new approach 
was applied to the simulate thermonuclear supernova explosions 
of \mch white dwarfs in 2 and 3 dimensions.  

An implicit assumption of our numerical model is that on the resolved scales
the flows are turbulent allowing us to describe the physics on the
unresolved scales by a subgrid scale model, in the spirit of large-eddy
simulations. The results presented here indicate 
that for supernova simulations this assumption is satisfied if we
compute 3D models and use at least a 256$^3$ grid. 
The reason is simply that more structure on small length scales 
in better resolved models increases the rate of fuel consumption 
locally, but the turbulent velocity fluctuations are then smaller 
on the grid scale, compensating for this gain.

All models we have computed (differing only in the ignition conditions
and the  grid resolution) explode. The explosion energy and the
Ni-masses are only moderately dependent on the way the nuclear 
flame is ignited making the explosions robust. However, since 
ignition is a stochastic process, the differences we find may 
even explain some of the spread in observed SN Ia's.

Based on our models we can predict lightcurves, spectra, and
abundances, and the first preliminary results look promising.
The lightcurves seem to be in very good agreement with observations, 
and also the nuclear abundances of elements and their
isotopes are found to be in the expected range. 
Of course, the next step is to compute a grid of models, with
varying white dwarf properties, and to compare them with the 
increasing data base of well-observed type Ia supernovae. 
The hope is that this will give us a tool to understand their 
physics and, thus, get confidence in their use for cosmology.

\section{Acknowledgements}

The numerical computations presented here were in part carried out on the 
Hitachi SR-8000 at the Leibniz-Rechenzentrum M\"unchen as a part 
of the project H007Z. We thank the staff of the HLRB at the LRZ for 
their continous help.  

This work was also supported in part by the Deutsche 
Forschungsgemeinschaft under Grant Hi 534/3-3.

\end{document}